# A Group-Theoretical Approach to the Periodic Table of Chemical Elements: Old and New Developments

Maurice R. Kibler

Institut de Physique Nucléaire
IN2P3-CNRS et Université Claude Bernard Lyon-1
43 Boulevard du 11 Novembre 1918
69622 Villeurbanne Cedex
France

**Abstract**

This paper is a companion article to the review paper by the present author devoted to the classification of matter constituents (chemical elements and particles) and published in the first part of the proceedings of The Second Harry Wiener International Memorial Conference. It is mainly concerned with a group-theoretical approach to the Periodic Table of the neutral elements based on the noncompact group SO(4, 2)⊗SU(2).

# A Group-Theoretical Approach to the Periodic Table of Chemical Elements: Old and New Developments

Maurice R. Kibler

Institut de Physique Nucléaire, IN2P3-CNRS et Université Claude Bernard Lyon-1, 43 Boulevard du 11 Novembre 1918, 69622 Villeurbanne Cedex, France

## 1. INTRODUCTION

This chapter is a companion article to the review paper by the present author devoted to the classification of matter constituents (chemical elements and particles) and published in the first part of the proceedings of The Second Harry Wiener International Memorial Conference [1]. It is mainly concerned with a group-theoretical approach to the Periodic Table of the neutral elements based on the group SO(4, 2)⊗SU(2).

The chapter is organized in the following way. The basic elements of group theory useful for this work are collected in Section 2. Section 3 deals with the Lie algebra of the Lie group SO(4, 2). In Section 4, the *à la* SO(4, 2)⊗SU(2) Periodic Table is presented and its potential interest is discussed in the framework of a program research (the KGR program).

## 2. BASIC GROUP THEORY

### 2.1 Some groups useful for the Periodic Table

The structure of a group is the simplest mathematical structure used in chemistry and physics. A group is a set of elements endowed with an internal composition law which is associative, has a neutral element, and for which each element in the set has an inverse with respect to the neutral element. There are two kinds of groups: the discrete groups (with a finite number or a countable infinite number of elements) and the continuous groups (with a noncountable infinite number of elements). Physics and chemistry use both kinds of groups. However, the groups relevant for the Periodic Table are continuous groups and more specifically Lie groups. Roughly speaking, a Lie group is a continuous group for which the composition law exhibits analyticity property. To a given Lie group corresponds one and only one Lie algebra (i.e., a nonassociative algebra such that the algebra law is anti-symmetric and satisfies the Jacobi identity). The reverse is not true in the sense that several Lie groups may correspond to a given Lie algebra.

As a typical example of a Lie group, let us consider the special real orthogonal group in three dimensions, noted SO(3), which is isomorphic to the point rotation group in three dimensions. The Lie algebra,[1] noted so(3) or $A_1$, of the Lie group SO(3) is nothing but the algebra of angular momentum of quantum mechanics. Indeed, the Lie algebra so(3) is isomorphic to the Lie algebra su(2) of the special unitary group in two dimensions SU(2). Therefore, the groups SO(3) and SU(2) have the same Lie algebra $A_1$. In more mathematical terms, the latter statement can be reformulated in three (equivalent) ways: su(2) is isomorphic to so(3),[2] or SU(2) is homomorphic onto SO(3) with a kernel of type $Z_2$, or SO(3) is isomorphic to SU(2)/$Z_2$.[3]

Among the Lie groups, we may distinguish: the simple Lie groups which do not have invariant Lie subgroup and the semi-simple Lie groups which do not have abelian (i.e., commutative) invariant Lie subgroup. This definition induces a corresponding definition for Lie algebras: we have simple Lie algebras (with no invariant Lie subalgebra) and semi-simple Lie algebras (with non abelian invariant Lie subalgebra). Of course, the Lie algebra of a semi-simple (respectively simple) Lie group is a semi-simple (respectively simple) Lie algebra. It should be noted that any semi-simple Lie algebra is a direct sum of simple Lie algebras.

For example, the Poincaré group (which is a Lie group leaving the form $dx^2+dy^2+dz^2–c^2dt^2$ invariant) is not a semi-simple Lie group. This group contains the space-time translations, the space spherical rotations and the Lorentz hyperbolic rotations. As a matter of fact, the space-time translations form an abelian invariant Lie subgroup so that the Poincaré group and its Lie algebra are neither semi-simple nor simple. On the contrary, the Lie groups SU(2) and SO(3) do not contain invariant Lie subgroup: they are simple and, consequently, semi-simple Lie groups. The Lie algebra su(2) ~ so(3) or $A_1$ is thus semi-simple.

An important point for applications to chemistry and physics is the Cartan classification of semi-simple Lie algebras. Let us consider a semi-simple Lie algebra A of dimension or order r (r is the dimension of the vector space associated with A). Let us note ( , ) the Lie algebra law (which is often a commutator [ , ] in numerous applications). It is possible to find a basis for the vector space associated with the algebra A, the so-called Cartan basis, the elements of which are noted $H_i$ with $i = 1, 2, …, \ell$ and $E_{\pm\alpha}$ with $\alpha = 1, 2, …, ½(r – \ell)$ and are such that

$$(H_i , H_j) = 0$$
$$(H_i , E_{+\alpha}) = \alpha_i E_{+\alpha}, \quad (H_i , E_{-\alpha}) = – \alpha_i E_{-\alpha}, \quad \alpha_i \in \mathbf{R}$$
$$(E_{+\alpha},E_{-\alpha}) = \alpha^i H_i, \quad (E_\alpha,E_\beta) = N_{\alpha\beta,\gamma} E_\gamma \text{ if } \beta \neq – \alpha, \quad \alpha^i \in \mathbf{R}, \quad N_{\alpha\beta,\gamma} \in \mathbf{R}$$

[1] The Lie algebra of a Lie group G is very often noted g. Hence, so(3) stands for the Lie algebra of the Lie group SO(3).
[2] This is noted su(2) ~ so(3).
[3] This is noted SO(3) ~ SU(2)/$Z_2$.

The Cartan basis clearly exhibits a non-invariant abelian Lie subalgebra, namely, the so-called Cartan subalgebra spanned by $H_1, H_2, \ldots, H_\ell$. The number $\ell$ is called the rank of the Lie algebra A. Of course, $r - \ell$ is necessarily even. Cartan showed that the semi-simple Lie algebras can be classified into:

- four families (each having a countable infinite number of members) noted $A_\ell$, $B_\ell$, $C_\ell$ and $D_\ell$
- five algebras referred to as exceptional algebras and noted $G_2$, $F_4$, $E_6$, $E_7$ and $E_8$.

The order r and the rank $\ell$ are indicated in the following tables for the four families and the five exceptional algebras. In addition, a representative (in real form) algebra is given for each family.

| **Family** | **Rank** $\ell$ | **Order** r | **Value of** $r - \ell$ | **Representative algebra** |
|---|---|---|---|---|
| $A_\ell$ | $\ell \geq 1$ | $\ell(\ell + 2)$ | $\ell(\ell + 1)$ | $su(\ell + 1)$ |
| $B_\ell$ | $\ell \geq 2$ | $\ell(2\ell + 1)$ | $2\ell^2$ | $so(2\ell + 1)$ |
| $C_\ell$ | $\ell \geq 3$ | $\ell(2\ell + 1)$ | $2\ell^2$ | $sp(2\ell)$ |
| $D_\ell$ | $\ell \geq 4$ | $\ell(2\ell - 1)$ | $2\ell(\ell - 1)$ | $so(2\ell)$ |

| **Exceptional algebra** | **Rank** $\ell$ | **Order** r | **Value of** $r - \ell$ |
|---|---|---|---|
| $G_2$ | 2 | 14 | 12 |
| $F_4$ | 4 | 52 | 48 |
| $E_6$ | 6 | 78 | 72 |
| $E_7$ | 7 | 133 | 126 |
| $E_8$ | 8 | 240 | 232 |

For each of the four families, a typical algebra is given:[4]

- the algebra su(n) with $n = \ell + 1$ for $A_\ell$, the Lie algebra of the special unitary group in n dimensions SU(n)
- the algebra so(n) with $n = 2\ell + 1$ for $B_\ell$ or $n = 2l$ for $D_\ell$, the Lie algebra of the special real orthogonal group in n dimensions SO(n)
- the algebra sp(2n) with $n = \ell$ for $C_\ell$, the Lie algebra of the symplectic group Sp(2n) in 2n dimensions.

Of course, other Lie algebras may occur for each family. For example, the Lie algebra so(4, 2) of the noncompact group SO(4, 2) is of type $D_3$, like the Lie algebra so(6) of the compact group SO(6).

[4] Some coincidences between a few families may appear for low values of $\ell$. For instance, we have $C_1 \equiv B_1 \equiv A_1$, $C_2 \equiv B_2$, $D_2 \equiv A_1 \oplus A_1$, and $D_3 \equiv A_3$.

A list of the Lie groups and Lie algebras to be used here is given in the following table. For each couple (group, algebra), the order r is the dimension of the corresponding Lie algebra; it is also the number of essential parameters necessary for characterizing each element of the corresponding Lie group.

| **Lie group** | SO(3) | SO(4) | SO(4, 1) | SO(3, 2) | SO(4, 2) | SU(2) | SU(2, 2) | Sp(8, **R**) |
|---|---|---|---|---|---|---|---|---|
| **Lie algebra** | $A_1$ | $D_2$ | $B_2$ | $B_2$ | $D_3$ | $A_1$ | $A_3$ | $C_4$ |
| **Order r** | 3 | 6 | 10 | 10 | 15 | 3 | 15 | 36 |

A word on each type of group in the table above is in order:

- The group SO(n) is the special real orthogonal group in n dimensions (n should not be confused with the order r = ½n(n–1) of the group). It is a compact group leaving invariant the real form $x_1^2 + x_2^2 + \ldots + x_n^2$ (with $x_i \in \mathbf{R}$) and is thus isomorphic to a point rotation group in $\mathbf{R}^n$.

- The group SO(p, q) is the special real pseudo-orthogonal group in p + q dimensions (p + q should not be confused with the dimension ½(p+q)(p+q–1) of the group). It is a noncompact group leaving invariant the real form $x_1^2 + x_2^2 + \ldots + x_p^2 - x_{p+1}^2 - x_{p+2}^2 - \ldots - x_{p+q}^2$ (with $x_i \in \mathbf{R}$) and is thus isomorphic to a generalized rotation group (involving spherical and hyperbolic rotations) in $\mathbf{R}^{p,q}$. Its maximal compact subgroup is SO(p)⊗SO(q). A well-known example is provided by SO(3, 1). The group SO(3, 1) is isomorphic to the Lorentz group, a sugroup of the Poincaré group, of order r = 6 spanned by 3 space rotations (around the axes x, y and z) and 3 space-time rotations (in the planes xt, yt and zt). Other examples are the groups SO(4, 1) and SO(3, 2), isomorphic to the two de Sitter groups which can be contracted (in the Wigner-Inönü sense) into the Poincaré group, and the group SO(4, 2), isomorphic to the conformal group. The Poincaré group itself is a subgroup of SO(4, 2).

- The group SU(n) is the special unitary group in n dimensions (n should not be confused with the dimension $r = n^2 - 1$ of the group). It is a compact group leaving invariant the Hermite complex form $|z_1|^2 + |z_2|^2 + \ldots + |z_p|^2$ (with $z_i \in \mathbf{C}$). The case n=2 corresponds to the spectral group that labels the spin since the Lie algebra of SU(2) is isomorphic to the algebra of a generalized angular momentum. The group SU(2) is the universal covering[5] group of SO(3), another way to say that SO(3) ~ SU(2)/$Z_2$. Note that the direct product SU(2)⊗SU(2) is homomorphic onto SO(4) with a kernel of type $Z_2$ so that we have the isomorphism SO(4) ~ SU(2)⊗SU(2)/$Z_2$ and SU(2)⊗SU(2) is the universal covering group of SO(4).[6]

---

[5] Given a Lie algebra A, there exists a simply connected Lie group whose Lie algebra is isomorphic to A. This group is the universal covering group of A.

[6] In terms of Lie algebras this means that so(4) ~ su(2) ⊕ su(2) or $D_2 \sim A_1 \oplus A_1$. Therefore, $D_2$ is a semi-simple Lie algebra which is not simple.

- The group SU(p, q) is the special pseudo-unitary group in p+q dimensions (p+q should not be confused with the dimension $r = (p+q)^2 - 1$ of the group). It is a noncompact group leaving invariant the complex form $|z_1|^2 + |z_2|^2 + \ldots + |z_p|^2 - |z_{p+1}|^2 - |z_{p+2}|^2 - \ldots - |z_{p+q}|^2$ (with $z_i \in \mathbf{C}$). The two groups SU(2, 2) and SO(4, 2) have the same Lie algebra $D_3$. The group SU(2, 2) is homomorphic onto SO(4, 2) with a kernel of type $Z_2$ so that SO(4, 2) ~ SU(2, 2)/$Z_2$ and thus SU(2, 2) is the universal covering group of SO(4, 2).

- The group Sp(2n, **R**) is the real symplectic group in 2n dimensions (2n should not be confused with the dimension r = n(2n+1) of the group). It is a noncompact group leaving invariant a symplectic form in 2n dimensions.

Most of the groups discussed above occur in the two chains of groups

$SO(3) \subset SO(3, 2) \subset SO(4, 2) \subset Sp(8, \mathbf{R})$
and
$SO(3) \subset SO(4) \subset SO(4, 1) \subset SO(4, 2) \subset Sp(8, \mathbf{R})$

which are of relevance in the present work.

**2.2 Representations of Lie groups and Lie algebras**

The concept of the representation of a group is essential for applications. A linear representation of dimension m of a group G is a homomorphic image of G in GL(m, **C**), the group of regular complex m × m matrices. In other words, to each element of G there corresponds a matrix of GL(m, **C**) such that the composition rules for G and GL(m, **C**) are conserved by this correspondence. Among the various representations of a group, the unitary irreducible representations play a central role. From the mathematical viewpoint, a unitary irreducible representation (UIR) consists of unitary matrices that leave no space for the representation invariant. If invariant subspaces exist, the representation is said to be reducible. From the practical viewpoint, a UIR corresponds to a set of states (wavefunctions in the framework of quantum mechanics) which can be exchanged or transformed into each other via the elements of the groups. More intuitively, we may think of a UIR of dimension m as a set of m boxes into each of which it is possible to put an object or state, the m objects in the set presenting analogies or similar properties. When passing from a group to one of its subgroups, each UIR of the group is generally reducible as a sum of UIR's of the subgroup.

A compact group has a countable infinite number of UIR's of finite dimension; the group SO(3) has UIR's of dimensions $2\ell + 1$, denoted $(\ell)$, where $\ell \in \mathbf{N}$, while the group SU(2) has UIR's of dimensions $2j + 1$, denoted (j), where $2j \in \mathbf{N}$; the group SO(4) has UIR's denoted $(j_1, j_2)$ with $2j_1 \in \mathbf{N}$ and $2j_2 \in \mathbf{N}$ of dimension $(2j_1 + 1)(2j_2 + 1)$; the UIR (j, j) of SO(4) can be decomposed into the direct sum $(j, j) = (0) \oplus (1) \oplus \ldots \oplus (2j)$ in terms of UIR's of SO(3). For a noncompact group, the UIR's are necessarily of infinite dimension; thus, each UIR of the noncompact group SO(4, 2) contains an infinite number of boxes.

The passage from a Lie group G to its corresponding Lie algebra g can be achieved by means of a derivation (or Taylor) process. It is thus clear that any representation of a Lie group provides a representation of its Lie algebra. The reverse passage from a Lie algebra g to one Lie group $G_i$ (the index i can take several values) is less evident since it involves an integration process. Therefore, from a given representation of a Lie algebra g, it is possible to generate either an ordinary (or uni-valued) representation of $G_i$ or a projective (or multi-valued) representation of $G_i$. The universal covering group G of all the groups Gi is the group for which any representation of g provides an ordinary representation of G. This is better understood with the example of the Lie algebra g = $A_1$ for which we have $G_1$ = SO(3) and $G_2$ = SU(2). The UIR's of $A_1$ are labeled (j) with $2j \in \mathbf{N}$, the UIR (j) being of dimension 2j + 1. In the case where j is an integer, the representation (j) of $A_1$ yields an ordinary representation for both SO(3) and SU(2) while in the case where j is half of an odd integer, the representation (j) of $A_1$ yields an ordinary representation of SU(2) and a projective (or spinor) representation of SO(3).[7]

**2.3 Representation theory of a semi-simple Lie algebra**

### 2.3.1 Generalities

In what follows, the elements of a Lie algebra A are linear operators X, Y, … acting on a separable Hilbert space and the Lie algebra law ( , ) is the commutator law [ , ] (with [X,Y] = XY – YX).[8] In this case (that corresponds to the general situation in quantum chemistry), we can construct polynomials from the elements of A and the algebra of these polynomials is called the enveloping algebra of A. Some particular polynomials play an important rôle, viz., those polynomials which commute with all the elements of A. These polynomials are called invariants of A. If A is semi-simple, the number of invariants of A and some of their properties are given by the three following theorems.

**Theorem 1 (Cartan).** The number of independent invariants of a semi-simple Lie algebra is equal to its rank.

Any semi-simple Lie algebra A of rank ℓ thus possesses ℓ invariant operators $C_i$ (with i = 1, 2, …, ℓ). One of these invariants, say $C_1$, is a polynomial of order 2 referred to as the Casimir or Casimir operator. Note that the remaining ℓ – 1 invariants of A are often called Casimir operators too.

---

[7] In the (Bethe) terminology of crystal-field theory familiar to the chemist, the group SU(2) is said to be the double group or spinor group of SO(3).

[8] Any basis of the vector space associated with the Lie algebra g = A gives a set of generators of its universal covering group G and each element of G can be obtained from the generators by means of an exponentiation process.

**Theorem 2 (Cartan).** For a semi-simple Lie algebra of rank $\ell$, the $\ell$ Cartan generators $H_i$ and the $\ell$ invariants $C_i$ with $i = 1, 2, \ldots, \ell$ constitute a set of commuting operators.

As a typical example, let us consider the algebra $A = A_1$ of order $r = 3$ and rank $\ell = 1$. Such an algebra is spanned by three operators $X_a$ ($a = 1, 2, 3$) that satisfy

$$[X_a, X_b] = \varepsilon_{abc} X_c$$

where $\varepsilon$ is the totally antisymmetric tensor on three indices with $\varepsilon_{123} = 1$. The Cartan basis ($H_1$, $E_{+1}$, $E_{-1}$) is given by

$$H_1 = (1/v)\, X_3, \quad E_{+1} = (1/v^2)\,(X_1 + iX_2), \quad E_{-1} = (1/v^2)\,(X_1 - iX_2), \quad v = \sqrt{2}$$

The Casimir operator is

$$C_1 = \tfrac{1}{2}\,(X_1^2 + X_2^2 + X_3^2) = H_1^2 + E_{+1}E_{-1} + E_{-1}E_{+1}$$

and the set $\{C_1, X_3\}$ is a set of commuting operators. Note that, in the theory of angular momentum, the latter set corresponds to the set $\{J^2, J_z\}$ of the square $J^2$ of a generalized angular momentum and its component $J_z$ around an arbitrary axis.

Going back to the general case of a semi-simple Lie algebra of order r and rank $\ell$, the set $\{H_i, C_i : i = 1, 2, \ldots, \ell\}$ is not generally a complete set of commuting operators. This can be made precise via Theorem 3.

**Theorem 3 (Racah).** For a semi-simple Lie algebra A of rank $\ell$ and order r, a further set of $f = \tfrac{1}{2}(r - 3\ell)$ additional operators in the enveloping algebra of A is necessary for completing the set $\{H_i, C_i : i = 1, 2, \ldots, \ell\}$.

Let $O_\beta$ (with $\beta = 1, 2, \ldots, f$) be the f additional operators. Then, Theorem 3 means that the set $\{H_i, C_i : i = 1, 2, \ldots, \ell;\ O_\beta : \beta = 1, 2, \ldots, f\}$ is a complete set consisting of $\ell + \ell + f = \tfrac{1}{2}(r + \ell)$ commuting operators. The values r, $\ell$ and f for some groups of interest here are given in the following table. (The number f is called the Racah number.)

| **Lie group** | SO(3) and SU(2) | SO(4) and SU(2)⊗SU(2) | SO(4, 1) and SO(3, 2) | SO(4, 2) and SU(2, 2) | Sp(8, **R**) |
|---|---|---|---|---|---|
| **Lie algebra** | $B_1 = A_1$ | $D_2 = A_1 \oplus A_1$ | $B_2$ | $D_3 = A_3$ | $C_4$ |
| **Order** r | 3 | 6 | 10 | 15 | 36 |
| **Rank** $\ell$ | 1 | 2 | 2 | 3 | 4 |
| **Racah number** f | 0 | 0 | 2 | 3 | 12 |
| **Number of commuting operators** | 2 | 4 | 6 | 9 | 20 |

For $A_1$, we have f = 0 and the set $\{C_1, X_3\}$ is complete, a well-known fact in the theory of angular momentum. As a consequence, we also have f = 0 for $D_2$. The situation is more complicated for the other algebras and we note that nine operators (3 $H_i$'s, 3 $C_i$'s and 3 $O_\beta$'s) occur for the algebra of SO(4, 2).

### 2.3.2 Representations of a semi-simple Lie algebra

Let H be the representation space of a semi-simple Lie algebra or order r and rank ℓ. The operators of the set $\{C_i, H_i : i = 1, 2, \ldots, \ell \,;\, O_\beta : \beta = 1, 2, \ldots, f\}$ admit common eigenvectors. We note $\psi((c_1, c_2, \ldots, c_\ell) \,;\, h_1, h_2, \ldots h_\ell \,;\, o_1, o_2, \ldots, o_f)$ or, in an abbreviated form, $\psi((c) \,;\, h \,;\, o)$ the common eigenvector of the ½ (r + ℓ) operators $C_i$, $H_i$ and $O_\beta$. The quantum numbers $c_1, c_2, \ldots, c_\ell$ stands for or characterize the eigenvalues of $C_1, C_2, \ldots, C_\ell$, respectively; similarly, $h_1, h_2, \ldots h_\ell$ and $o_1, o_2, \ldots, o_f$ stand for or characterize the eigenvalues of $H_1, H_2, \ldots, H_\ell$ and $O_1, O_2, \ldots, O_f$, respectively. More precisely, we have

$$C_i\, \psi((c) \,;\, h \,;\, o) = c_i\, \psi((c) \,;\, h \,;\, o)$$
$$H_i\, \psi((c) \,;\, h \,;\, o) = h_i\, \psi((c) \,;\, h \,;\, o)$$
$$O_\beta\, \psi((c) \,;\, h \,;\, o) = o_\beta\, \psi((c) \,;\, h \,;\, o)$$

For fixed $(c) \equiv (c_1, c_2, \ldots, c_\ell)$, the vectors $\psi((c) \,;\, h \,;\, o)$, with $h \equiv h_1, h_2, \ldots, h_\ell$ and $o \equiv o_1, o_2, \ldots, o_f$ varying, span an irreducible representation of the Lie algebra A. In other words, the eigenvalues $c_1, c_2, \ldots, c_\ell$ (or the numbers characterizing the eigenvalues) of the invariants $C_1, C_2, \ldots, C_\ell$ may serve for labeling the irreducible representations of A.

## 3. THE LIE ALGEBRA SO(4, 2) AS A DYNAMICAL NON-INVARIANCE ALGEBRA

### 3.1 The chain SO(3) ⊂ SO(4)

Let

$$H\, \Psi = E \Psi$$

be the Schrödinger equation for a hydrogen-like atom of nuclear charge Ze (Z = 1 for hydrogen). We recall that the discrete spectrum (E < 0) of the non-relativistic Hamiltonian H is given by

$$E \equiv E_n = E_1/n^2, \quad \Psi \equiv \Psi_{n\ell m}$$

where $E_1$ is the energy of the ground state (depending on the reduced mass of the hydrogen-like system, the nuclear charge and the Planck constant) and where

$n = 1, 2, 3, \ldots$
for fixed $n : \ell = 0, 1, \ldots, n - 1$
for fixed $\ell : m = -\ell, -\ell + 1, \ldots, \ell$

The degeneracy degree of the level $E_n$ is $n^2$ if the spin of the electron is not taken into account ($2n^2$ if taken into account).[9]

For fixed n and ℓ, the degeneracy of the 2ℓ+1 wavefunctions $\Psi_{n\ell m}$ is explained by the three-dimensional proper rotation group, isomorphic to SO(3). The degeneracy in m follows from the existence of a first constant of the motion, i.e., the angular momentum **L**(L1,L2,L3) of the electron of the hydrogen-like atom. Indeed, we have a first set of three commuting operators: the Hamiltonian H, the square $\mathbf{L}^2$ of the angular momentum of the electron and the third component $L_3$ of the angular momentum. The operators $L_1$, $L_2$ and $L_3$ span the Lie algebra of SO(3).

For fixed n, the degeneracy of the wavefunctions $\Psi_{n\ell m}$ corresponding to different values of ℓ is accidental with respect to SO(3) and can be explained via the group SO(4). The introduction of SO(4) can be achieved in the framework either of a local approach by Pauli [2] or a stereographic[10] approach by Fock [3]. We shall follow here the local or Lie-like approach developed by Pauli [2]. The degeneracy in ℓ follows from the existence of a second constant of the motion, i.e., the Runge-Lenz vector $\mathbf{M}(M_1,M_2,M_3)$ (discussed by Laplace, Hamilton, Runge and Lenz in classical mechanics and by Pauli in quantum mechanics). By properly rescaling the vector **M**, we obtain a vector $\mathbf{A}(A_1,A_2,A_3)$ which has the dimension of an angular momentum. It can be shown that the set $\{L_i, A_i : i = 1, 2, 3\}$ spans the Lie algebra of SO(4), SO(3, 1) and E(3) for $E < 0$, $E > 0$ and $E = 0$, respectively.[11] As a point of fact, the obtained Lie algebras are Lie algebras with constraints since there exists two quadratic relations between **L** and **A** [2]. We now continue with the case $E < 0$. Then, by defining

$$\mathbf{J} = \mathbf{L} + \mathbf{A}, \quad \mathbf{K} = \mathbf{L} - \mathbf{A}$$

it is possible to write the Lie algebra of SO(4) in the form of the Lie algebra of the direct product $SO(3)_J \otimes SO(3)_K$ with the sets

$$\{J_i = L_i + A_i : i = 1, 2, 3\}$$

and

$$\{K_i = L_i - A_i : i = 1, 2, 3\}$$

generating $SO(3)_J$ and $SO(3)_K$, respectively. The UIR's of SO(4) can be labeled as (j, k) with $2j \in \mathbf{N}$ and $2k \in \mathbf{N}$. For fixed n, the $n^2$ functions $\Psi_{n\ell m}$ generate the UIR (½ (n – 1), ½ (n – 1)) of SO(4), of dimension $n^2$, corresponding to $2j = 2k = n - 1$.

[9] Note that $2n^2$ can take the values $2n^2 = 2, 8, 18, 32, 50$, etc.

[10] Fock mapped the Schrödinger equation for the hydrogen atom in $\mathbf{R}^3$ onto the one for a spherically symmetrical free-point rotor in $\mathbf{R}^4$ (in momentum space after Fourier transform). This stereographic mapping establishes a connection between the motion of a particle in a Coulomb potential in the three-dimensional space ($\mathbf{R}^3$ hydrogen atom) and a particle constrained to move in a null potential on the surface of a four-dimensional sphere ($\mathbf{R}^4$ free rotor) [3].

[11] E(3) stands for the Euclidean group in three dimensions.

The condition j = k comes from one of the two quadratic relations between **L** and **A** (which can be transcribed as $\mathbf{J}^2 = \mathbf{K}^2$ in terms of **J** and **K**). The other relation gives $E = E_1/n^2$. Finally, note that the restriction from SO(4) to SO(3) yields the decomposition

$$(½ (n – 1), ½ (n – 1)) = (0) \oplus (1) \oplus \ldots \oplus (n – 1)$$

in terms of UIR's (ℓ) of SO(3).

At this point, it is interesting to understand the different status of the groups SO(3) and SO(4) for an hydrogen-like atom. The group SO(4), respectively SO(3), describes multiplets characterized by a given value of the principal quantum number n, respectively the orbital angular quantum number ℓ; the multiplet of SO(4), respectively SO(3), associated to n, respectively ℓ, is of dimension $n^2$, respectively 2ℓ+1. The group SO(3) is a geometrical symmetry group in the sense that it leaves invariant both the kinetic part and the potential part of the Hamiltonian H. As a consequence, the generators $L_1$, $L_2$ and $L_3$ of SO(3) commute with H. Hence, we have the set $\{H, \mathbf{L}^2, L_3\}$ of commuting operators. The group SO(4) is a dynamical invariance group that manifests itself here via its Lie algebra: it does not correspond to a geometrical symmetry group of H but its generators commute with H. Hence, we have the set $\{H, \mathbf{L}^2, L_3, \mathbf{A}^2, A_3\}$ of commuting operators. The latter set comprises five independent constants of the motion (see Ref. [4]).[12]

Besides the symmetry group SO(3) and the dynamical invariance group SO(4), another group, namely SO(4, 2) plays an important rôle. This is a dynamical non-invariance group in the sense that not all the generators of SO(4, 2) commute with H. It can be considered as a spectrum generating group since any wavefunction $\Psi_{n\ell m}$ can be deduced from any other wavefunction through the action on $\Psi_{n\ell m}$ of generators of SO(4, 2) which do not commute with H. The group SO(4, 2) was introduced, in connection with the hydrogen atom, by Barut and Kleinert [5] and independently, in a SO(6, **C**) form, by Malkin and Man'ko [6]. We now give the general lines for constructing SO(4, 2) from SO(4) by an ascent process (see Refs. [5], [7] and [8]).

### 3.2 The chain SO(3) ⊂ SO(4) ⊂ SO(4, 2)

The starting point is to find a bosonic realization of the Lie algebra of SO(4). Let us introduce two commuting pairs of boson operators $(a_1, a_2)$ and $(a_3, a_4)$. They satisfy the commutation relations

$$[a_i, a_j^+] = \delta(i, j), \quad [a_i, a_j] = [a_i^+, a_j^+] = 0$$

[12] This number (2d – 1 = 5) is the maximum number of independent constants of the motion for a dynamical system in d = 3 dimensions. A system in d dimensions for which the number of independent constants of the motion is equal to 2d – 1 is referred to a maximally superintegrable system [4]. For d = 3, two maximally superintegrable systems are known: the harmonic oscillator system and the hydrogen-like system.

where we use $A^+$ to denote the Hermitean conjugate of the operator A. (The $a_i$'s are annihilation operators and the $a_i^+$'s are creation operators.) It is a simple matter of calculation to check that the six bilinear forms ($J_{12}$, $J_{23}$, $J_{31}$) and ($J_{14}$, $J_{24}$, $J_{34}$) defined via

$$J_{\alpha\beta} = \tfrac{1}{2} (a^+ \sigma_\gamma a + b^+ \sigma_\gamma b) \text{ with } \alpha, \beta, \gamma \text{ cyclic}$$

and

$$J_{\alpha 4} = -\tfrac{1}{2} (a^+ \sigma_\alpha a - b^+ \sigma_\alpha b)$$

span the Lie algebra $D_2$ of SO(4). In the latter two definitions, we have

- the indices $\alpha$, $\beta$, $\gamma$ can take the values 1, 2 and 3
- $\sigma_1$, $\sigma_2$ and $\sigma_3$ are the Pauli matrices
- a and b stand for the column vectors whose transposed vectors are the line vectors ${}^t a = (a_1\ a_2)$ and ${}^t b = (a_3\ a_4)$, respectively
- $a^+$ and $b^+$ stand for the line vectors $a^+ = (a_1^+\ a_2^+)$ and $b^+ = (a_3^+\ a_4^+)$, respectively.

We can find nine additional operators which together with the two triplets of operators ($J_{12}$, $J_{23}$, $J_{31}$) and ($J_{14}$, $J_{24}$, $J_{34}$) generate the Lie algebra $D_3$ of SO(4, 2). These generators can be defined as

$$J_{\alpha 5} = i[J_{\alpha 4}, J_{45}] \quad \text{with } \alpha = 1, 2, 3$$
$$J_{45} = \tfrac{1}{2} (a^+ \sigma_2\, {}^t b^+ - a\, \sigma_2\, {}^t b)$$
$$J_{\alpha 6} = - i[J_{\alpha 5}, J_{56}] \quad \text{with } \alpha = 1, 2, 3$$
$$J_{46} = - i[J_{45}, J_{56}]$$
$$J_{56} = \tfrac{1}{2} (a^+ a + b^+ b + 2)$$

The fifteen operators ($J_{12}$, $J_{23}$, $J_{31}$), ($J_{14}$, $J_{24}$, $J_{34}$), on one hand, and ($J_{15}$, $J_{25}$, $J_{35}$, $J_{45}$), ($J_{16}$, $J_{26}$, $J_{36}$, $J_{46}$, $J_{56}$), on the other hand, can be shown to satisfy

$$[J_{ab}, J_{cd}] = i(g_{bc}J_{ad} - g_{ac}J_{bd} + g_{ad}J_{bc} - g_{bd}J_{ac})$$

where the metric ($g_{ab}$) is defined by

$$(g_{ab}) = \mathrm{diag}(-1 -1 -1 -1\, 1\, 1)$$

We thus end up with the Lie algebra of SO(4, 2). A set of infinitesimal generators of SO(4, 2), acting on functions $f : (x_1, x_2, \ldots, x_6) \mapsto f(x_1, x_2, \ldots, x_6)$ of six variables, is provided by the differential operators

$$J_{ab} \mapsto U_{ab} = i(g_{aa} x_a \partial/\partial x_b - g_{bb} x_b \partial/\partial x_a)$$

which generalize the components of the usual orbital angular momentum in $\mathbf{R}^3$.[13]

Since the Lie algebra of SO(4, 2) is of rank 3, we have three invariant operators in its enveloping algebra. Indeed, they are of degree 2, 3 and 4 and are given by

$$C_1 = \sum J_{ab} J^{ab}$$
$$C_2 = \sum \varepsilon_{abcdef} J^{ab} J^{cd} J^{ef}$$
$$C_3 = \sum J_{ab} J^{bc} J_{cd} J^{da}$$

where $\varepsilon$ is the totally anti-symmetric tensor on six covariant indexes with $\varepsilon_{123456} = 1$ and

$$J^{ab} = \sum g^{ac} g^{bd} J_{cd} \quad \text{with } g^{ab} = g_{ab}$$

(we use the Einstein summation conventions).

We now briefly discuss the existence of a special representation of SO(4, 2) that provides the quantum numbers n, $\ell$ and m. It can be seen that the operator $J_{45}$ connects states having different values of n. This shows that the Lie algebra of the group SO(4, 2) is a dynamical non-invariance algebra. In fact, it is possible to construct any state vector $\psi_{n\ell m}$ by repeated application on the ground state vector $\psi_{100}$ of the operator $J_{45} - J_{46}$ and of shift operators for SO(4). This result is at the root of the fact that all discrete levels (or bound states) of an hydrogen-like atom span an UIR of SO(4, 2). Let h be this representation. It corresponds to the eigenvalues 6, 0 and –12 of the invariant operators $C_1$, $C_2$ and $C_3$, respectively. The dynamical non-invariance group SO(4, 2) contains another dynamical non-invariance group, namely SO(4, 1), that contains in turn the dynamical invariance group SO(4) and thus the symmetry group SO(3). The representation h remains irreducible when SO(4, 2) is restricted to SO(4, 1). On the other side, the restriction of SO(4, 2) to SO(4) yields the decomposition

$$h = \oplus_{2j \in \mathbf{N}} (j, j)$$

which is a direct sum of the representations (j, j), with $2j \in \mathbf{N}$, of the group SO(4). Further restriction from SO(4) to SO(3) gives

$$h = \oplus_{n \in \mathbf{N}^*} \oplus_{\ell = 0 \text{ to } n-1} (\ell)$$

that reflects that all the discrete states (with $E < 0$) of a hydrogen-like atom are contained in a single UIR of SO(4, 2). A similar result holds for the continuum states (with $E > 0$).[14]

---

[13] The expressions of the commutator $[J_{ab}, J_{cd}]$ and of the infinitesimal generator $U_{ab}$ are also valid for SO(p, q) if we take $(g_{ab}) = (-1-1\ldots-111..1)$ where –1 occurs p times and +1 occurs q times.

### 3.3 The chain SO(4, 2) ⊂ Sp(8, R)

There exists a connection between the Schrödinger equation for a three-dimensional hydrogen-like atom and the one for a four-dimensional isotropic harmonic oscillator.[15] Such a connection can be derived from the $\mathbf{R}^4 \to \mathbf{R}^3$ Kustaanheimo-Stiefel transformation, which is nothing but the Hopf fibration $S^3\times\mathbf{R}^+ \to S^2\times\mathbf{R}^+$ of compact fiber $S^1$ (see Refs. [9, 10, 11] and references therein). This connection can also be obtained from a bosonization, via Jordan-Schwinger boson calculus, of the Pauli equations for a hydrogen-like atom: the four-dimensional harmonic oscillator can thus be decomposed into a pair of coupled two-dimensional isotropic oscillators with the same energy [9]. This coupled pair of $\mathbf{R}^2$ oscillators, or alternatively the $\mathbf{R}^4$ oscillator, can be described by the real symplectic group in eight dimensions Sp(8, **R**).[16] The introduction in the Lie algebra of Sp(8, **R**) of the constraint arising from the equality of the energy of the two two-dimensional oscillators yields a Lie algebra under constraints[17] which is isomorphic to the Lie algebra of SU(2, 2), the pseudo-unitary group in 2+2 dimensions, a covering group of SO(4, 2). The dynamical non-invariance group SO(4, 2) for a three-dimensional hydrogen-like atom thus arises as a quotient of the real symplectic group associated to a four-dimensional isotropic harmonic oscillator subjected to a constraint. This derivation of SO(4, 2) corresponds to a symmetry descent process and should be contrasted with the one in Section 3.2 which corresponds to a symmetry ascent process. A link between the two derivations can be established by noting that the Lie algebra of Sp(8, **R**) can be generated by the 36 possible bilinear forms of $a_i$ and $a_i^+$ (i = 1 to 4) among which 15 generate the Lie algebra of SU(2, 2).

## 4. AN SO(4, 2)⊗SU(2) APPROACH TO THE PERIODIC TABLE

### 4.1 Why the group SO(4, 2)⊗SU(2)?

There are several group-theoretical approaches to the periodic table. We mainly deal here with the SO(4, 2)⊗SU(2) approach based on the group SO(4, 2) [12, 13, 14, 15, 16, 17, 18, 19]. An interesting question is to ask why the group SO(4, 2) is a quite natural candidate for a group-theoretical description of the periodic table. As a first reason, chemistry is determined by the electromagnetic interaction (one of the four fundamental interactions besides the weak, strong and gravitational interactions) and we know from the beginning of the twentieth century that the Maxwell equations of the electromagnetic field are covariant under the conformal group in the Minkowski space, a group isomorphic to SO(4, 2) [20].[18] The second reason comes from the fact that SO(4, 2) is the maximal dynamical non-invariance group for a hydrogen-like atom which accounts for all the usual spectroscopic quantum numbers used for complex atoms described in a central-field approximation. Of course, SO(4, 2) contains the groups SO(4) and SO(3), the latter group being the geometrical symmetry group for any complex atom.[19] Finally, the group SU(2) that enters in SO(4, 2)⊗SU(2) is a spectral group which makes it possible to double the number of states described by SO(4, 2).

---

[14] Let us remember that the analog of SO(4) for the continuum (or scattering) states of a hydrogen-like atom is the noncompact dynamical invariance group SO(3, 1), another subgroup of SO(4, 2).

[15] The oscillator is an attractive (i.e., ordinary) oscillator for $E < 0$, a repulsive oscillator for $E > 0$. It reduces to a free-particle system for $E = 0$.

[16] The real symplectic group Sp(2N, **R**) in 2N dimensions is a dynamical non-invariance group for the isotropic harmonic oscillator in N dimensions. Its maximal compact subgroup SU(N) is a dynamical invariance group for the oscillator.

[17] The concept of Lie algebra under constraints was discussed from the physical point of view in Ref. [9] and defined in the mathematical sense in Ref. [11].

To be complete, we should also mention other group-theoretical approaches to the Periodic Table [21, 22, 23, 24]. Most of the other approaches are based on a joint use of quantum mechanics and group theory both applied to atomic spectroscopy. Among these approaches, we may cite the one by Novaro and Wolf [21] based on perturbation theory and the one by Négadi and Kibler [24] based on quantum groups and deformation theory. Along these lines, let us mention that the SU(2)⊗SU(2)⊗SU(2) approach by Novaro and Berrondo [21] is in some sense connected to the SO(4, 2)⊗SU(2) approach to be described below since the direct product SU(2)⊗SU(2)⊗SU(2) occurs in the noncanonical chain SO(4, 2)⊗SU(2) ⊃ SO(4)⊗SU(2) via the isomorphism SO(4) ~ SU(2)⊗SU(2)/$Z_2$.

### 4.2 The *à la* SO(4, 2)⊗SU(2) periodic table

We discuss here the Periodic System of the chemical elements following the general lines developed by Barut [13] on the basis of the group SO(4, 2) and by Byakov, Kulakov, Rumer and Fet [16] on the basis of the (direct product) group SO(4, 2)⊗SU(2) and further investigated by Odabasi [14], Kibler [18], and Gurskii et al. [19] (see also the pioneer works in Refs. [12, 15, 17]).

From a practical point of view, there is one representation of SO(4, 2), viz., the representation h described in Sections 2 and 3, that contains all discrete states of the hydrogen atom: The set of all discrete states of the hydrogen atom can be regarded as spanning an infinite multiplet of SO(4, 2). This is the starting point for the construction of a Periodic Chart based on SO(4, 2)⊗SU(2). In such a construction, the chemical elements can be treated as structureless particles and/or considered as the various possible states of a dynamical system; each state can be deduced from a reference state (a vacuum state in high energy physics terminology) by means of shift (or creation) operators.

---

[18] The group SO(4, 2) is the largest group leaving Maxwell's equations invariant.

[19] For a complex atom (many-electron atom), the SO(4) invariance is no longer valid and only the SO(3) symmetry remains. The SO(4) invariance is broken by an SO(3) symmetry breaking term (as is the case in an Hartree-Fock treatment of a many-electron atom).

To start off, let us give a two-dimensional graphical representation of the UIR h of the group SO(4, 2). The infinite multiplet h can be organized into multiplets of SO(4) and SO(3). The multiplets of SO(4) may be placed into rows characterized by the UIR labels n = 1, 2, etc. The $n^{th}$ row contains n multiplets of SO(3) characterized by the UIR labels $\ell = 0, 1, \ldots, n-1$. This leads to a frame, with rows labeled by n and columns by $\ell$, of entries $[n+\ell\ n]$ located at line n and column $\ell$, in complete analogy with the skeleton described for the Madelung rule [18]. With the entry $[n+\ell\ n]$ are associated $2\ell + 1$ boxes and each box can be divided into two boxes so that the entry $[n+\ell\ n]$ contains $2(2\ell + 1)$ boxes. The resultant doubling, i.e., $2\ell + 1 \to 2(2\ell + 1)$, may be accounted for by the introduction of the group SU(2): We thus pass from $SO(4, 2) \supset SO(4) \supset SO(3)$ to $SO(4, 2)\otimes SU(2) \supset SO(4)\otimes SU(2) \supset SO(3)\otimes SU(2)$. The $2(2\ell + 1)$ boxes in the entry $[n+\ell\ n]$ are organized into multiplets (j) of the group SU(2): For $\ell$ different from 0, the entry contains two multiplets corresponding to $j = \ell - \frac{1}{2}$ and $j = \ell + \frac{1}{2}$ of lengths $2\ell$ and $2(\ell + \frac{1}{2}) + 1 = 2(\ell + 1)$, respectively; for $\ell = 0$, the entry contains only one multiplet corresponding to $j = \frac{1}{2}$ of length 2. Each box in the entry $[n+\ell\ n]$ of the frame can be characterized by an address $(n\ell jm)$ with j and, for fixed j, m increasing from left to right (for fixed j, the values of m are $m = -j, -j + 1, \ldots, j$). We thus obtain a Chart [18] for which the $n^{th}$ row contains $2n^2$ (with $2n^2 = 2, 8, 18, 32, 50$, etc.) boxes and the $\ell^{th}$ column contains an infinite number of boxes.

The connection with chemistry is as follows. To each address $(n\ell jm)$ we can associate a value of the atomic number Z (see Section 4.3). The box $(n\ell jm)$ is then filled with the chemical element of atomic number Z. This produces Table 1. It is remarkable that Table 1 is very close to the one arising from the Madelung rule [18]. The only difference is that in Table 1 the elements in a given entry $[n+\ell\ n]$ are arranged in one or two multiplets according to whether $\ell$ is zero or different from zero. The rows n = 1, 2, 3, etc. of Table 1 correspond to the shells K, L, M, etc., respectively, encountered in the standard Periodic Table. A given column of Table 1 corresponds to a family of chemical analogs in the standard Periodic Table.

Table 1 resembles, to some extent, other Periodic Tables. First, Table 1 is, up to the exchange of rows and columns, identical to the Table introduced by Byakov, Kulakov, Rumer and Fet [16]. The presentation adopted here, which is based on former work by the author, leads to a Table the format of which is easily comparable to that of most Tables in present day use. Second, Table 1 exhibits, up to a rearrangement, the same blocks as the table by Neubert [25] based on the filling of only four Coulomb shells. Third, Table 1 presents, up to a rearrangement, some similarities with the table by Dash [26] based on the principal quantum number, the law of second-order constant energy differences and the Coulomb-momentum interaction, except that in the Dash table the third transition group does not begin with Lu (Z = 71), the lanthanide series does not run from La (Z = 57) to Yb (Z = 70) and the actinide series does not run from Ac (Z = 89) to No (Z = 102). Finally, it can be seen that Table 1 is very similar to the Periodic Table presented by Scerri at the Second Harry Wiener International Conference [27]. As a point of fact, to pass from the format of Table 1 to the format of Scerri's Table, it is sufficient to perform a symmetry operation with respect to an axis parallel to the columns and located at the left of Table 1 and then to get down the p-blocks (for ℓ = 1) by one unit, the d-blocks (for ℓ = 2) by two units, etc.

The main distinguishing features of Table 1 are seen to be the following: (i) hydrogen is in the family of the alkali metals, (ii) helium belongs to the family of the alkaline earth metals, and (iii) the inner transition series (lanthanides and actinides) as well as the transition series (iron group, palladium group and platinum group) occupy a natural place in the table. This contrasts with the conventional Tables with 8(9) or 18 columns where: (i) hydrogen is sometimes located in the family of the halogens, (ii) helium generally belongs to the family of the noble gases, and (iii) the lanthanide series and the actinide series are generally treated as appendages. Furthermore, in Table 1 the distribution of the elements Z = 104 to 120 is in agreement with Seaborg's predictions based on atomic shell calculations. In contrast to his predictions, however, Table 1 shows that the elements Z = 121 to 138 form a new family having no homologue among the known elements. In addition, Table 1 suggests that the family of super-actinides contains the elements Z = 139 to 152 (and not Z = 122 to 153 as predicted by Seaborg).

We have mentioned in Table 1 the name darmstadtium recently given to the element Z = 110 and mentioned the recent observation of the elements Z = 113 and Z = 115 [28]. The next table gives some information concerning some of the elements discovered in the last four decades.

| **Z** | 105 | 106 | 107 | 108 | 109 | 110 |
|---|---|---|---|---|---|---|
| **Element** | dubnium | seaborgium | bohrium | hassium | meitnerium | darmstadtium |
| **Symbol** | Db | Sg | Bh | Hs | Mt | Ds |
| **Year** | 1968 | 1974 | 1981 | 1984 | 1982 | 1994 |

| **Z** | 111 | 112 | 113 | 114 | 115 | 116 |
|---|---|---|---|---|---|---|
| **Element** | not named | not named | not named | not named | not named | not named |
| **Symbol** | / | / | / | | / | / |
| **Year** | 1994 | 1996 | 2004 | 1999 | 2004 | 1999 |

Finally, note that the number of elements afforded by Table 1 is a priori infinite. The hunting of superheavy elements is not closed.

Another important peculiarity of Table 1 is the division, for ℓ different from zero, of the ℓ-block into two sub-blocks of length 2ℓ and 2(ℓ+1). As an illustration, we get two subblocks of length 4 and 6 for the d-blocks (corresponding to ℓ = 2) and two subblocks of length 6 and 8 for the f-blocks (corresponding to ℓ = 3). In the case of rare earth elements, the division into light or ceric rare earth elements and heavy or yttric rare earth elements is quite well known. It has been underlined on the basis of differences for the solubilities of some mineral salts and on the basis of physical properties as, for example, magnetic and spectroscopic properties of rare earth compounds [29]. The division for the rare earth elements and the other ℓ-blocks (with $\ell$ different from 0) was explained by using the Dirac theory for the electron. In particular, it was shown by Oudet [29] that the calculation of magnetic moments via this theory accounts for the division of the rare earth elements. It is to be noted that the valence 2 of the samarium ion can be explained by the division of the f-block into two subblocks. In a shell-model picture, the ion $Sm^{2+}$ then corresponds to a completely filled shell (with six electrons) associated with the first subblock.

### 4.3 Atomic number of an element

There is a one-to-one correspondence between chemical elements and state vectors of the space for the representation h of SO(4, 2). Therefore, each atom in Table 1 can be characterized by an address which depends on the labeling of the state vectors. This address can be written ($n\ell jm$) in terms of the quantum numbers arising from the groups SO(4), SO(3) and SU(2). From the address ($n\ell jm$) of a neutral element, we can obtain its atomic number $Z(n\ell jm)$ owing to the formula:

$$Z(n\ell jm) = (n + \ell)[(n + \ell)^2 - 1]/6 + (n + \ell + 1)^2/2 - [1 + (-1)^{n+\ell}](n + \ell + 1)/4 \\ - 4\,\ell(\ell + 1) + \ell + j(2\ell + 1) + m - 1$$

Table 2 lists the various chemical elements in terms of the quantum numbers ($n\ell jm$). The sum $n+\ell$, which occurs in the Madelung rule [30], plays an important rôle in this formula. The formula can be specialized to two specific formulas according to whether $n+\ell$ is even or odd. In this respect, it should be noted that the states with $n+\ell$ even, on one side, and the states with $n+\ell$ odd, on the other side, span two distinct UIR's, say $h_e$ and $h_o$, respectively, of the noncompact subgroup SO(3, 2) of SO(4, 2). The UIR h of SO(4, 2) can be decomposed as

$$h = h_o \oplus h_e$$

when restricting SO(4, 2) to SO(3, 2). The neutral elements can thus be separated into two classes according to whether as $n+\ell$ is even or odd. The classification into two classes is in agreement with experimental data: the ionization potentials for the chemical elements can be described by two curves, one for $n+\ell$ odd and the other for $n+\ell$ even. This clearly emphasizes the relevance of the chain of groups

$$SO(4, 2) \supset SO(3, 2) \supset SO(3) \otimes SO(2)$$

introduced by Barut [13] for neutral elements. This chain has to be distinguished from the chain of groups

$$SO(4, 2) \supset SO(4, 1) \supset SO(4) \supset SO(3)$$

which corresponds to the hydrogenic order and is thus more adapted to highly ionized atoms [13].

By using shift operators (raising and lowering operators) for the chain

$$SO(4, 2)\otimes SU(2) \supset SO(4)\otimes SU(2) \supset SO(3)\otimes SU(2) \supset SO(2)\otimes SU(2),$$

it is possible to connect elements with addresses having the same value of n (for example, elements of the lanthanide series). In a similar way, the use of shift operators for the chain

$$SO(4, 2)\otimes SU(2) \supset SO(2)\otimes SO(2, 1)\otimes SU(2) \supset SO(2)\otimes SO(2)\otimes SU(2)$$

allows to connect elements with addresses having the same value of ℓ (for example, elements of the family of the alkaline earth metals). It is even possible to find operators which render possible the knight's move in the Periodic Table introduced by Laing [31]. More generally, the address of an arbitrary element can be deduced from the address of hydrogen by means of repeated actions of shift operators. The shift operators make it possible to connect one vector of the space of the UIR h to one another, i.e., to pass from one chemical element to another. In the SO(4, 2) approach to the Periodic Table, the neutral atoms are partners which can be transformed one into the other under the action of the generators of SO(4, 2). A similar result applies to ionized atoms.

### 4.4 Electron configuration of an element

Given the address (nℓjm) of a neutral element, it is also possible to obtain its atomic configuration if we have a filling order for the atomic shells. As an example, let us take the Madelung rule for ordering the shells. This rule leads to the following order of shell filling (see also Ref. [32] ):[20]

$$1s < 2s < 2p < 3s < 3p < 4s < 3d < 4p < 5s < 4d < 5p < 6s < 4f < 5d < 6p < 7s < 5f < 6d < 7p < 8s < \dots \quad (1)$$

In order to find the electron configuration of the element characterized by the address (nℓjm), we isolate in (1) the nℓ shell. Then, each νλ shell preceding the nℓ shell is filled with 2(2ν+1) electrons, according to the Stoner prescription. Hence, $Z_c$ electrons are used for the completely filled shells and we put the $Z - Z_c$ remaining electrons on the nℓ shell. This completes the derivation of the electron configuration of the element $Z \equiv Z(n\ell jm)$.

[20] The Madelung rule for neutral elements, noted [n+ℓ, n]↑, corresponds to an ordering with n+ℓ increasing and, for n+ℓ fixed, with n increasing. This rule is violated by some twenty exceptions between hydrogen (Z = 1) to einsteinium (Z = 99) [32]. There are of course several other rules for ordering the shells. The most well-known are: the [n–½ℓ, n]↑ rule connected with the harmonic oscillator, the [n, ℓ]↑ rule connected with the hydrogen atom and the [n+½ℓ, n]↑ [32]. There also exists some intermediate rules which can be derived by varying a free parameter originating from perturbation or deformation theory [21, 24].

We give in Table 3 the electron configurations derived from spectroscopic data and/or self-consistent field calculations. The results can be compared to the ones derived from the Madelung rule.

Let us close this Section with some remarks about the Madelung rule. First, the sequence (1), ordered according to n+ℓ increasing, corresponds to a maximal filling with 2, 2, 8, 8, 18, 18, 32, 32, … electrons for n+ℓ = 1, 2, 3, 4, 5, 6, 7, 8, …, respectively. Second, the Madelung rule (that is inherent to the Aufbau Prinzip of Bohr [33]) corresponds to an SO(4) → SO(3) symmetry breaking where the SO(4) levels

1s ⊕ {2s ⊕ 2p} ⊕ {3s ⊕ 3p ⊕ 3d} ⊕ {4s ⊕ 4p ⊕ 4d ⊕ 4f} ⊕ {5s ⊕ 5p ⊕…

are split into the SO(3) levels

1s ⊕ 2s ⊕ 2p ⊕ 3s ⊕ 3p ⊕ 4s ⊕ 3d ⊕ 4p ⊕ 5s ⊕ 4d ⊕ 5p ⊕ 6s ⊕ 4f ⊕ …

in increasing energy. (The levels between brackets have the same energy.) This symmetry breaking is not to be taken in the sense of high energy physics (Higgs mechanism) since it is not spontaneous but rather due to electron correlation.

### 4.5 The KGR program

The application of the group SO(4, 2)⊗SU(2) has up to now been limited to qualitative aspects of the Periodic Table. We would now like to present in outline a program (inherited from nuclear and particle physics) for using SO(4, 2)⊗SU(2) in a quantitative way. A detailed treatment will be presented in subsequent papers.

We have seen in Sections 2 and 3 that the group SO(4, 2), locally isomorphic to the special unitary group in 2 + 2 dimensions SU(2, 2), is a Lie group of order 15 and rank 3. It has therefore 15 generators involving three Cartan generators (i.e., generators commuting between themselves). In addition, it has three invariants or Casimir operators (i.e., independent polynomials in the generators that commute with all generators of the group). Therefore, we have a set of 3 + 3 = 6 operators that commute between themselves. Indeed, this set is not complete from the mathematical point of view. According to the lemma by Racah, we need to find ½(order – 3 rank) = 3 additional operators in order to get a complete set. It is thus possible to build a complete set of 6 + 3 = 9 commuting operators. Each of the nine operators can be taken to be self-adjoint and thus, from the quantum-mechanical point of view, can describe an observable.

The next step is to connect chemical and physical properties (like ionization energies, electron affinities, electronegativities, melting and boiling points, specific heats, atomic radii, atomic volumes, densities, magnetic susceptibilities, solubilities, etc.) to the nine commuting operators. In most cases, this can be done by expressing a chemical observable associated with a given property as a linear combination of the nine commuting operators.

The last step is to fit the various linear combinations to experimental data. For each property this will lead to a formula or phenomenological law that can be used in turn for making predictions concerning the chemical elements for which no data are available.

This program, referred to as the KGR program since it was discussed at the Kananaskis Guest Ranch during the 2003 Harry Wiener International Conference, is ambitious. Its realization needs the collaboration of chemists, physicists and mathematicians. Persons interested in this program may write to the author.

| | | | | | | | | | | | | | | | | | | | | | | | | | | | | | | | | |
|---|---|---|---|---|---|---|---|---|---|---|---|---|---|---|---|---|---|---|---|---|---|---|---|---|---|---|---|---|---|---|---|---|
| 1 | 2 | | | | | | | | | | | | | | | | | | | | | | | | | | | | | | | |
| H | He | | | | | | | | | | | | | | | | | | | | | | | | | | | | | | | |
| 3 | 4 | 5 | 6 | 7 | 8 | 9 | 10 | | | | | | | | | | | | | | | | | | | | | | | | | |
| Li | Be | B | C | N | O | F | Ne | | | | | | | | | | | | | | | | | | | | | | | | | |
| 11 | 12 | 13 | 14 | 15 | 16 | 17 | 18 | 21 | 22 | 23 | 24 | 25 | 26 | 27 | 28 | 29 | 30 | | | | | | | | | | | | | | | |
| Na | Mg | Al | Si | P | S | Cl | Ar | Sc | Ti | V | Cr | Mn | Fe | Co | Ni | Cu | Zn | | | | | | | | | | | | | | | |
| 19 | 20 | 31 | 32 | 33 | 34 | 35 | 36 | 39 | 40 | 41 | 42 | 43 | 44 | 45 | 46 | 47 | 48 | 57 | 58 | 59 | 60 | 61 | 62 | 63 | 64 | 65 | 66 | 67 | 68 | 69 | 70 | |
| K | Ca | Ga | Ge | As | Se | Br | Kr | Y | Zr | Nb | Mo | Tc | Ru | Rh | Pd | Ag | Cd | La | Ce | Pr | Nd | Pm | Sm | Eu | Gd | Tb | Dy | Ho | Er | Tm | Yb | |
| 37 | 38 | 49 | 50 | 51 | 52 | 53 | 54 | 71 | 72 | 73 | 74 | 75 | 76 | 77 | 78 | 79 | 80 | 89 | 90 | 91 | 92 | 93 | 94 | 95 | 96 | 97 | 98 | 99 | 100 | 101 | 102 | 121-138 |
| Rb | Sr | In | Sn | Sb | Te | I | Xe | Lu | Hf | Ta | W | Re | Os | Ir | Pt | Au | Hg | Ac | Th | Pa | U | Np | Pu | Am | Cm | Bk | Cf | Es | Fm | Md | No | no |
| 55 | 56 | 81 | 82 | 83 | 84 | 85 | 86 | 103 | 104 | 105 | 106 | 107 | 108 | 109 | 110 | 111 | 112 | 139 | 140 | 141 | 142 | 143 | 144 | 145 | 146 | 147 | 148 | 149 | 150 | 151 | 152 | … |
| Cs | Ba | Tl | Pb | Bi | Po | At | Rn | Lr | Rf | Db | Sg | Bh | Hs | Mt | Ds | X? | X? | no | no | no | no | no | no | no | no | no | no | no | no | no | no | … |
| 87 | 88 | 113 | 114 | 115 | 116 | 117 | 118 | 153 | 154 | 155 | 156 | 157 | 158 | 159 | 160 | 161 | 162 | … | | | | | | | | | | | | | | |
| Fr | Ra | X? | X? | X? | X? | no | no | no | no | no | no | no | no | no | no | no | no | … | | | | | | | | | | | | | | |
| … | | | | | | | | | | | | | | | | | | | | | | | | | | | | | | | | |
| … | | | | | | | | | | | | | | | | | | | | | | | | | | | | | | | | |

**Table 1: The *à la* SO(4, 2)⊗SU(2) Periodic Table (X? = observed but not named, no = not observed)**

**Table 2: Addresses (nljm) of the known chemical elements**

| n | l | j | m | [n+l n] | Z | Element | Symbol |
|---|---|---|---|---|---|---|---|
| 1 | 0 | 0.5 | -0.5 | 1 1 | 1 | hydrogen | H |
| 1 | 0 | 0.5 | 0.5 | 1 1 | 2 | helium | He |
| 2 | 0 | 0.5 | -0.5 | 2 2 | 3 | lithium | Li |
| 2 | 0 | 0.5 | 0.5 | 2 2 | 4 | berylium | Be |
| 2 | 1 | 0.5 | -0.5 | 3 2 | 5 | boron | B |
| 2 | 1 | 0.5 | 0.5 | 3 2 | 6 | carbon | C |
| 2 | 1 | 1.5 | -1.5 | 3 2 | 7 | nitrogen | N |
| 2 | 1 | 1.5 | -0.5 | 3 2 | 8 | oxygen | O |
| 2 | 1 | 1.5 | 0.5 | 3 2 | 9 | fluorine | F |
| 2 | 1 | 1.5 | 1.5 | 3 2 | 10 | **neon** | **Ne** |
| | | | | | | | |
| 3 | 0 | 0.5 | -0.5 | 3 3 | 11 | sodium | Na |
| 3 | 0 | 0.5 | 0.5 | 3 3 | 12 | magnesium | Mg |
| 3 | 1 | 0.5 | -0.5 | 4 3 | 13 | aluminium | Al |
| 3 | 1 | 0.5 | 0.5 | 4 3 | 14 | silicon | Si |
| 3 | 1 | 1.5 | -1.5 | 4 3 | 15 | phosphorous | P |
| 3 | 1 | 1.5 | -0.5 | 4 3 | 16 | sulfur | S |
| 3 | 1 | 1.5 | 0.5 | 4 3 | 17 | chlorine | Cl |
| 3 | 1 | 1.5 | 1.5 | 4 3 | 18 | **argon** | **Ar** |
| | | | | | | | |
| 4 | 0 | 0.5 | -0.5 | 4 4 | 19 | potassium | K |
| 4 | 0 | 0.5 | 0.5 | 4 4 | 20 | calcium | Ca |
| 3 | 2 | 1.5 | -1.5 | 5 3 | 21 | scandium | Sc |
| 3 | 2 | 1.5 | -0.5 | 5 3 | 22 | titanium | Ti |
| 3 | 2 | 1.5 | 0.5 | 5 3 | 23 | vanadium | V |
| 3 | 2 | 1.5 | 1.5 | 5 3 | 24 | chromium | Cr |
| 3 | 2 | 2.5 | -2.5 | 5 3 | 25 | manganese | Mn |
| 3 | 2 | 2.5 | -1.5 | 5 3 | 26 | iron | Fe |
| 3 | 2 | 2.5 | -0.5 | 5 3 | 27 | cobalt | Co |
| 3 | 2 | 2.5 | 0.5 | 5 3 | 28 | nickel | Ni |
| 3 | 2 | 2.5 | 1.5 | 5 3 | 29 | copper | Cu |
| 3 | 2 | 2.5 | 2.5 | 5 3 | 30 | zinc | Zn |
| 4 | 1 | 0.5 | -0.5 | 5 4 | 31 | gallium | Ga |
| 4 | 1 | 0.5 | 0.5 | 5 4 | 32 | germanium | Ge |
| 4 | 1 | 1.5 | -1.5 | 5 4 | 33 | arsenic | As |
| 4 | 1 | 1.5 | -0.5 | 5 4 | 34 | selenium | Se |
| 4 | 1 | 1.5 | 0.5 | 5 4 | 35 | bromine | Br |
| 4 | 1 | 1.5 | 1.5 | 5 4 | 36 | **krypton** | **Kr** |

**Table 2 (continued)**

| n | l | j | m | [n+l n] | Z | Element | Symbol |
|---|---|---|---|---|---|---|---|
| 5 | 0 | 0.5 | -0.5 | 5 5 | 37 | rubidium | Rb |
| 5 | 0 | 0.5 | 0.5 | 5 5 | 38 | strontium | Sr |
| 4 | 2 | 1.5 | -1.5 | 6 4 | 39 | yttrium | Y |
| 4 | 2 | 1.5 | -0.5 | 6 4 | 40 | zirconium | Zr |
| 4 | 2 | 1.5 | 0.5 | 6 4 | 41 | niobium | Nb |
| 4 | 2 | 1.5 | 1.5 | 6 4 | 42 | molybdenum | Mo |
| 4 | 2 | 2.5 | -2.5 | 6 4 | 43 | technetium | Tc |
| 4 | 2 | 2.5 | -1.5 | 6 4 | 44 | ruthenium | Ru |
| 4 | 2 | 2.5 | -0.5 | 6 4 | 45 | rhodium | Rh |
| 4 | 2 | 2.5 | 0.5 | 6 4 | 46 | palladium | Pd |
| 4 | 2 | 2.5 | 1.5 | 6 4 | 47 | silver | Ag |
| 4 | 2 | 2.5 | 2.5 | 6 4 | 48 | cadmium | Cd |
| 5 | 1 | 0.5 | -0.5 | 6 5 | 49 | indium | In |
| 5 | 1 | 0.5 | 0.5 | 6 5 | 50 | tin | Sn |
| 5 | 1 | 1.5 | -1.5 | 6 5 | 51 | antimony | Sb |
| 5 | 1 | 1.5 | -0.5 | 6 5 | 52 | tellurium | Te |
| 5 | 1 | 1.5 | 0.5 | 6 5 | 53 | iodine | I |
| 5 | 1 | 1.5 | 1.5 | 6 5 | 54 | **xenon** | **Xe** |
| 6 | 0 | 0.5 | -0.5 | 6 6 | 55 | cesium | Cs |
| 6 | 0 | 0.5 | 0.5 | 6 6 | 56 | barium | Ba |
| 4 | 3 | 2.5 | -2.5 | 7 4 | 57 | lanthanum | La |
| 4 | 3 | 2.5 | -1.5 | 7 4 | 58 | cerium | Ce |
| 4 | 3 | 2.5 | -0.5 | 7 4 | 59 | praseodymium | Pr |
| 4 | 3 | 2.5 | 0.5 | 7 4 | 60 | neodymium | Nd |
| 4 | 3 | 2.5 | 1.5 | 7 4 | 61 | promethium | Pm |
| 4 | 3 | 2.5 | 2.5 | 7 4 | 62 | samarium | Sm |
| 4 | 3 | 3.5 | -3.5 | 7 4 | 63 | europium | Eu |
| 4 | 3 | 3.5 | -2.5 | 7 4 | 64 | gadolinium | Gd |
| 4 | 3 | 3.5 | -1.5 | 7 4 | 65 | terbium | Tb |
| 4 | 3 | 3.5 | -0.5 | 7 4 | 66 | dysprosium | Dy |
| 4 | 3 | 3.5 | 0.5 | 7 4 | 67 | holmium | Ho |
| 4 | 3 | 3.5 | 1.5 | 7 4 | 68 | erbium | Er |
| 4 | 3 | 3.5 | 2.5 | 7 4 | 69 | thulium | Tm |
| 4 | 3 | 3.5 | 3.5 | 7 4 | 70 | ytterbium | Yb |
| 5 | 2 | 1.5 | -1.5 | 7 5 | 71 | lutetium | Lu |
| 5 | 2 | 1.5 | -0.5 | 7 5 | 72 | hafnium | Hf |
| 5 | 2 | 1.5 | 0.5 | 7 5 | 73 | tantalum | Ta |
| 5 | 2 | 1.5 | 1.5 | 7 5 | 74 | tungsten | W |
| 5 | 2 | 2.5 | -2.5 | 7 5 | 75 | rhenium | Re |
| 5 | 2 | 2.5 | -1.5 | 7 5 | 76 | osmium | Os |

**Table 2 (continued)**

| n | l | j | m | [n+l n] | Z | Element | Symbol |
|---|---|---|---|---|---|---|---|
| 5 | 2 | 2.5 | -0.5 | 7 5 | 77 | iridium | Ir |
| 5 | 2 | 2.5 | 0.5 | 7 5 | 78 | platinum | Pt |
| 5 | 2 | 2.5 | 1.5 | 7 5 | 79 | gold | Au |
| 5 | 2 | 2.5 | 2.5 | 7 5 | 80 | mercury | Hg |
| 6 | 1 | 0.5 | -0.5 | 7 6 | 81 | thallium | Tl |
| 6 | 1 | 0.5 | 0.5 | 7 6 | 82 | lead | Pb |
| 6 | 1 | 1.5 | -1.5 | 7 6 | 83 | bismuth | Bi |
| 6 | 1 | 1.5 | -0.5 | 7 6 | 84 | polonium | Po |
| 6 | 1 | 1.5 | 0.5 | 7 6 | 85 | astatine | At |
| 6 | 1 | 1.5 | 1.5 | 7 6 | 86 | **radon** | **Rn** |
| 7 | 0 | 0.5 | -0.5 | 7 7 | 87 | francium | Fr |
| 7 | 0 | 0.5 | 0.5 | 7 7 | 88 | radium | Ra |
| 5 | 3 | 2.5 | -2.5 | 8 5 | 89 | actinium | Ac |
| 5 | 3 | 2.5 | -1.5 | 8 5 | 90 | thorium | Th |
| 5 | 3 | 2.5 | -0.5 | 8 5 | 91 | protactinium | Pa |
| 5 | 3 | 2.5 | 0.5 | 8 5 | 92 | uranium | U |
| 5 | 3 | 2.5 | 1.5 | 8 5 | 93 | neptunium | Np |
| 5 | 3 | 2.5 | 2.5 | 8 5 | 94 | plutonium | Pu |
| 5 | 3 | 3.5 | -3.5 | 8 5 | 95 | americium | Am |
| 5 | 3 | 3.5 | -2.5 | 8 5 | 96 | curium | Cm |
| 5 | 3 | 3.5 | -1.5 | 8 5 | 97 | berkelium | Bk |
| 5 | 3 | 3.5 | -0.5 | 8 5 | 98 | californium | Cf |
| 5 | 3 | 3.5 | 0.5 | 8 5 | 99 | einsteinium | Es |
| 5 | 3 | 3.5 | 1.5 | 8 5 | 100 | fermium | Fm |
| 5 | 3 | 3.5 | 2.5 | 8 5 | 101 | mendelevium | Md |
| 5 | 3 | 3.5 | 3.5 | 8 5 | 102 | nobelium | No |
| 6 | 2 | 1.5 | -1.5 | 8 6 | 103 | lawrencium | Lr |
| 6 | 2 | 1.5 | -0.5 | 8 6 | 104 | rutherfordium | Rf |
| 6 | 2 | 1.5 | 0.5 | 8 6 | 105 | dubnium | Db |
| 6 | 2 | 1.5 | 1.5 | 8 6 | 106 | seaborgium | Sg |
| 6 | 2 | 2.5 | -2.5 | 8 6 | 107 | bohrium | Bh |
| 6 | 2 | 2.5 | -1.5 | 8 6 | 108 | hassium | Hs |
| 6 | 2 | 2.5 | -0.5 | 8 6 | 109 | meitneirium | Mt |
| 6 | 2 | 2.5 | 0.5 | 8 6 | 110 | darmstadtium | Ds |
| 6 | 2 | 2.5 | 1.5 | 8 6 | 111 | not named | no symbol |
| 6 | 2 | 2.5 | 2.5 | 8 6 | 112 | not named | no symbol |
| 7 | 1 | 0.5 | -0.5 | 8 7 | 113 | not named | no symbol |
| 7 | 1 | 0.5 | 0.5 | 8 7 | 114 | not named | no symbol |
| 7 | 1 | 1.5 | -1.5 | 8 7 | 115 | not named | no symbol |
| 7 | 1 | 1.5 | -0.5 | 8 7 | 116 | not named | no symbol |

**Table 3: Electronic configurations for the elements from Z = 1 to Z = 103**

| Z | Element | | 1s | 2s | 2p | 3s | 3p | 4s | 3d | 4p | 5s | 4d | 5p | 6s | 4f | 5d | 6p | 7s | 5f | 6d |
|---|---|---|---|---|---|---|---|---|---|---|---|---|---|---|---|---|---|---|---|---|
| 1 | hydrogen | H | 1 | | | | | | | | | | | | | | | | | |
| 2 | helium | He | 2 | | | | | | | | | | | | | | | | | |
| 3 | lithium | Li | 2 | 1 | | | | | | | | | | | | | | | | |
| 4 | berylium | Be | 2 | 2 | | | | | | | | | | | | | | | | |
| 5 | boron | B | 2 | 2 | 1 | | | | | | | | | | | | | | | |
| 6 | carbon | C | 2 | 2 | 2 | | | | | | | | | | | | | | | |
| 7 | nitrogen | N | 2 | 2 | 3 | | | | | | | | | | | | | | | |
| 8 | oxygen | O | 2 | 2 | 4 | | | | | | | | | | | | | | | |
| 9 | fluorine | F | 2 | 2 | 5 | | | | | | | | | | | | | | | |
| 10 | **neon** | **Ne** | **2** | **2** | **6** | | | | | | | | | | | | | | | |
| 11 | sodium | Na | 2 | 2 | 6 | 1 | | | | | | | | | | | | | | |
| 12 | magnesium | Mg | 2 | 2 | 6 | 2 | | | | | | | | | | | | | | |
| 13 | aluminium | Al | 2 | 2 | 6 | 2 | 1 | | | | | | | | | | | | | |
| 14 | silicon | Si | 2 | 2 | 6 | 2 | 2 | | | | | | | | | | | | | |
| 15 | phosphorous | P | 2 | 2 | 6 | 2 | 3 | | | | | | | | | | | | | |
| 16 | sulfur | S | 2 | 2 | 6 | 2 | 4 | | | | | | | | | | | | | |
| 17 | chlorine | Cl | 2 | 2 | 6 | 2 | 5 | | | | | | | | | | | | | |
| 18 | **argon** | **Ar** | **2** | **2** | **6** | **2** | **6** | | | | | | | | | | | | | |
| 19 | potassium | K | 2 | 2 | 6 | 2 | 6 | 1 | | | | | | | | | | | | |
| 20 | calcium | Ca | 2 | 2 | 6 | 2 | 6 | 2 | | | | | | | | | | | | |
| 21 | scandium | Sc | 2 | 2 | 6 | 2 | 6 | 2 | 1 | | | | | | | | | | | |
| 22 | titanium | Ti | 2 | 2 | 6 | 2 | 6 | 2 | 2 | | | | | | | | | | | |
| 23 | vanadium | V | 2 | 2 | 6 | 2 | 6 | 2 | 3 | | | | | | | | | | | |
| 24 | chromium | Cr | 2 | 2 | 6 | 2 | 6 | 1 | 5 | | | | | | | | | | | |
| 25 | manganese | Mn | 2 | 2 | 6 | 2 | 6 | 2 | 5 | | | | | | | | | | | |
| 26 | iron | Fe | 2 | 2 | 6 | 2 | 6 | 2 | 6 | | | | | | | | | | | |
| 27 | cobalt | Co | 2 | 2 | 6 | 2 | 6 | 2 | 7 | | | | | | | | | | | |
| 28 | nickel | Ni | 2 | 2 | 6 | 2 | 6 | 2 | 8 | | | | | | | | | | | |
| 29 | copper | Cu | 2 | 2 | 6 | 2 | 6 | 1 | 10 | | | | | | | | | | | |
| 30 | zinc | Zn | 2 | 2 | 6 | 2 | 6 | 2 | 10 | | | | | | | | | | | |
| 31 | gallium | Ga | 2 | 2 | 6 | 2 | 6 | 2 | 10 | 1 | | | | | | | | | | |
| 32 | germanium | Ge | 2 | 2 | 6 | 2 | 6 | 2 | 10 | 2 | | | | | | | | | | |
| 33 | arsenic | As | 2 | 2 | 6 | 2 | 6 | 2 | 10 | 3 | | | | | | | | | | |
| 34 | selenium | Se | 2 | 2 | 6 | 2 | 6 | 2 | 10 | 4 | | | | | | | | | | |
| 35 | bromine | Br | 2 | 2 | 6 | 2 | 6 | 2 | 10 | 5 | | | | | | | | | | |
| 36 | **krypton** | **Kr** | **2** | **2** | **6** | **2** | **6** | **2** | **10** | **6** | | | | | | | | | | |
| 37 | rubidium | Rb | 2 | 2 | 6 | 2 | 6 | 2 | 10 | 6 | 1 | | | | | | | | | |
| 38 | strontium | Sr | 2 | 2 | 6 | 2 | 6 | 2 | 10 | 6 | 2 | | | | | | | | | |
| 39 | yttrium | Y | 2 | 2 | 6 | 2 | 6 | 2 | 10 | 6 | 2 | 1 | | | | | | | | |
| 40 | zirconium | Zr | 2 | 2 | 6 | 2 | 6 | 2 | 10 | 6 | 2 | 2 | | | | | | | | |
| 41 | niobium | Nb | 2 | 2 | 6 | 2 | 6 | 2 | 10 | 6 | 1 | 4 | | | | | | | | |
| 42 | molybdenum | Mo | 2 | 2 | 6 | 2 | 6 | 2 | 10 | 6 | 1 | 5 | | | | | | | | |
| 43 | technetium | Tc | 2 | 2 | 6 | 2 | 6 | 2 | 10 | 6 | 1 | 6 | | | | | | | | |
| 44 | ruthenium | Ru | 2 | 2 | 6 | 2 | 6 | 2 | 10 | 6 | 1 | 7 | | | | | | | | |
| 45 | rhodium | Rh | 2 | 2 | 6 | 2 | 6 | 2 | 10 | 6 | 1 | 8 | | | | | | | | |
| 46 | palladium | Pd | 2 | 2 | 6 | 2 | 6 | 2 | 10 | 6 | | 10 | | | | | | | | |
| 47 | silver | Ag | 2 | 2 | 6 | 2 | 6 | 2 | 10 | 6 | 1 | 10 | | | | | | | | |
| 48 | cadmium | Cd | 2 | 2 | 6 | 2 | 6 | 2 | 10 | 6 | 2 | 10 | | | | | | | | |
| 49 | indium | In | 2 | 2 | 6 | 2 | 6 | 2 | 10 | 6 | 2 | 10 | 1 | | | | | | | |
| 50 | tin | Sn | 2 | 2 | 6 | 2 | 6 | 2 | 10 | 6 | 2 | 10 | 2 | | | | | | | |
| 51 | antimony | Sb | 2 | 2 | 6 | 2 | 6 | 2 | 10 | 6 | 2 | 10 | 3 | | | | | | | |

**Table 3 (continued)**

| **Z** | **Element** | | **1s** | **2s** | **2p** | **3s** | **3p** | **4s** | **3d** | **4p** | **5s** | **4d** | **5p** | **6s** | **4f** | **5d** | **6p** | **7s** | **5f** | **6d** |
|---|---|---|---|---|---|---|---|---|---|---|---|---|---|---|---|---|---|---|---|---|
| 52 | tellurium | Te | 2 | 2 | 6 | 2 | 6 | 2 | 10 | 6 | 2 | 10 | 4 | | | | | | | |
| 53 | iodine | I | 2 | 2 | 6 | 2 | 6 | 2 | 10 | 6 | 2 | 10 | 5 | | | | | | | |
| 54 | **xenon** | **Xe** | **2** | **2** | **6** | **2** | **6** | **2** | **10** | **6** | **2** | **10** | **6** | | | | | | | |
| 55 | cesium | Cs | 2 | 2 | 6 | 2 | 6 | 2 | 10 | 6 | 2 | 10 | 6 | 1 | | | | | | |
| 56 | barium | Ba | 2 | 2 | 6 | 2 | 6 | 2 | 10 | 6 | 2 | 10 | 6 | 2 | | | | | | |
| 57 | lanthanum | La | 2 | 2 | 6 | 2 | 6 | 2 | 10 | 6 | 2 | 10 | 6 | 2 | | 1 | | | | |
| 58 | cerium | Ce | 2 | 2 | 6 | 2 | 6 | 2 | 10 | 6 | 2 | 10 | 6 | 2 | 2 | | | | | |
| 59 | praseodymium | Pr | 2 | 2 | 6 | 2 | 6 | 2 | 10 | 6 | 2 | 10 | 6 | 2 | 3 | | | | | |
| 60 | neodymium | Nd | 2 | 2 | 6 | 2 | 6 | 2 | 10 | 6 | 2 | 10 | 6 | 2 | 4 | | | | | |
| 61 | promethium | Pm | 2 | 2 | 6 | 2 | 6 | 2 | 10 | 6 | 2 | 10 | 6 | 2 | 5 | | | | | |
| 62 | samarium | Sm | 2 | 2 | 6 | 2 | 6 | 2 | 10 | 6 | 2 | 10 | 6 | 2 | 6 | | | | | |
| 63 | europium | Eu | 2 | 2 | 6 | 2 | 6 | 2 | 10 | 6 | 2 | 10 | 6 | 2 | 7 | | | | | |
| 64 | gadolinium | Gd | 2 | 2 | 6 | 2 | 6 | 2 | 10 | 6 | 2 | 10 | 6 | 2 | 7 | 1 | | | | |
| 65 | terbium | Tb | 2 | 2 | 6 | 2 | 6 | 2 | 10 | 6 | 2 | 10 | 6 | 2 | 9 | | | | | |
| 66 | dysprosium | Dy | 2 | 2 | 6 | 2 | 6 | 2 | 10 | 6 | 2 | 10 | 6 | 2 | 10 | | | | | |
| 67 | holmium | Ho | 2 | 2 | 6 | 2 | 6 | 2 | 10 | 6 | 2 | 10 | 6 | 2 | 11 | | | | | |
| 68 | erbium | Er | 2 | 2 | 6 | 2 | 6 | 2 | 10 | 6 | 2 | 10 | 6 | 2 | 12 | | | | | |
| 69 | thulium | Tm | 2 | 2 | 6 | 2 | 6 | 2 | 10 | 6 | 2 | 10 | 6 | 2 | 13 | | | | | |
| 70 | ytterbium | Yb | 2 | 2 | 6 | 2 | 6 | 2 | 10 | 6 | 2 | 10 | 6 | 2 | 14 | | | | | |
| 71 | lutecium | Lu | 2 | 2 | 6 | 2 | 6 | 2 | 10 | 6 | 2 | 10 | 6 | 2 | 14 | 1 | | | | |
| 72 | hafnium | Hf | 2 | 2 | 6 | 2 | 6 | 2 | 10 | 6 | 2 | 10 | 6 | 2 | 14 | 2 | | | | |
| 73 | tantalum | Ta | 2 | 2 | 6 | 2 | 6 | 2 | 10 | 6 | 2 | 10 | 6 | 2 | 14 | 3 | | | | |
| 74 | tungsten | W | 2 | 2 | 6 | 2 | 6 | 2 | 10 | 6 | 2 | 10 | 6 | 2 | 14 | 4 | | | | |
| 75 | rhenium | Re | 2 | 2 | 6 | 2 | 6 | 2 | 10 | 6 | 2 | 10 | 6 | 2 | 14 | 5 | | | | |
| 76 | osmium | Os | 2 | 2 | 6 | 2 | 6 | 2 | 10 | 6 | 2 | 10 | 6 | 2 | 14 | 6 | | | | |
| 77 | iridium | Ir | 2 | 2 | 6 | 2 | 6 | 2 | 10 | 6 | 2 | 10 | 6 | 2 | 14 | 7 | | | | |
| 78 | platinum | Pt | 2 | 2 | 6 | 2 | 6 | 2 | 10 | 6 | 2 | 10 | 6 | 1 | 14 | 9 | | | | |
| 79 | gold | Au | 2 | 2 | 6 | 2 | 6 | 2 | 10 | 6 | 2 | 10 | 6 | 1 | 14 | 10 | | | | |
| 80 | mercury | Hg | 2 | 2 | 6 | 2 | 6 | 2 | 10 | 6 | 2 | 10 | 6 | 2 | 14 | 10 | | | | |
| 81 | thallium | Tl | 2 | 2 | 6 | 2 | 6 | 2 | 10 | 6 | 2 | 10 | 6 | 2 | 14 | 10 | 1 | | | |
| 82 | lead | Pb | 2 | 2 | 6 | 2 | 6 | 2 | 10 | 6 | 2 | 10 | 6 | 2 | 14 | 10 | 2 | | | |
| 83 | bismuth | Bi | 2 | 2 | 6 | 2 | 6 | 2 | 10 | 6 | 2 | 10 | 6 | 2 | 14 | 10 | 3 | | | |
| 84 | polonium | Po | 2 | 2 | 6 | 2 | 6 | 2 | 10 | 6 | 2 | 10 | 6 | 2 | 14 | 10 | 4 | | | |
| 85 | astatine | At | 2 | 2 | 6 | 2 | 6 | 2 | 10 | 6 | 2 | 10 | 6 | 2 | 14 | 10 | 5 | | | |
| 86 | **radon** | **Rn** | **2** | **2** | **6** | **2** | **6** | **2** | **10** | **6** | **2** | **10** | **6** | **2** | **14** | **10** | **6** | | | |
| 87 | francium | Fr | 2 | 2 | 6 | 2 | 6 | 2 | 10 | 6 | 2 | 10 | 6 | 2 | 14 | 10 | 6 | 1 | | |
| 88 | radium | Ra | 2 | 2 | 6 | 2 | 6 | 2 | 10 | 6 | 2 | 10 | 6 | 2 | 14 | 10 | 6 | 2 | | |
| 89 | actinium | Ac | 2 | 2 | 6 | 2 | 6 | 2 | 10 | 6 | 2 | 10 | 6 | 2 | 14 | 10 | 6 | 2 | | 1 |
| 90 | thorium | Th | 2 | 2 | 6 | 2 | 6 | 2 | 10 | 6 | 2 | 10 | 6 | 2 | 14 | 10 | 6 | 2 | | 2 |
| 91 | protactinium | Pa | 2 | 2 | 6 | 2 | 6 | 2 | 10 | 6 | 2 | 10 | 6 | 2 | 14 | 10 | 6 | 2 | 2 | 1 |
| 92 | uranium | U | 2 | 2 | 6 | 2 | 6 | 2 | 10 | 6 | 2 | 10 | 6 | 2 | 14 | 10 | 6 | 2 | 3 | 1 |
| 93 | neptunium | Np | 2 | 2 | 6 | 2 | 6 | 2 | 10 | 6 | 2 | 10 | 6 | 2 | 14 | 10 | 6 | 2 | 4 | 1 |
| 94 | plutonium | Pu | 2 | 2 | 6 | 2 | 6 | 2 | 10 | 6 | 2 | 10 | 6 | 2 | 14 | 10 | 6 | 2 | 6 | |
| 95 | americium | Am | 2 | 2 | 6 | 2 | 6 | 2 | 10 | 6 | 2 | 10 | 6 | 2 | 14 | 10 | 6 | 2 | 7 | |
| 96 | curium | Cm | 2 | 2 | 6 | 2 | 6 | 2 | 10 | 6 | 2 | 10 | 6 | 2 | 14 | 10 | 6 | 2 | 7 | 1 |
| 97 | berkelium | Bk | 2 | 2 | 6 | 2 | 6 | 2 | 10 | 6 | 2 | 10 | 6 | 2 | 14 | 10 | 6 | 2 | 8 | 1 |
| 98 | californium | Cf | 2 | 2 | 6 | 2 | 6 | 2 | 10 | 6 | 2 | 10 | 6 | 2 | 14 | 10 | 6 | 2 | 10 | |
| 99 | einsteinium | Es | 2 | 2 | 6 | 2 | 6 | 2 | 10 | 6 | 2 | 10 | 6 | 2 | 14 | 10 | 6 | 2 | 11 | |
| 100 | fermium | Fm | 2 | 2 | 6 | 2 | 6 | 2 | 10 | 6 | 2 | 10 | 6 | 2 | 14 | 10 | 6 | 2 | 12 | |
| 101 | mendelevium | Md | 2 | 2 | 6 | 2 | 6 | 2 | 10 | 6 | 2 | 10 | 6 | 2 | 14 | 10 | 6 | 2 | 13 | |
| 102 | nobelium | No | 2 | 2 | 6 | 2 | 6 | 2 | 10 | 6 | 2 | 10 | 6 | 2 | 14 | 10 | 6 | 2 | 14 | |
| 103 | lawrencium | Lr | 2 | 2 | 6 | 2 | 6 | 2 | 10 | 6 | 2 | 10 | 6 | 2 | 14 | 10 | 6 | 2 | 14 | 1 |

## 5. CLOSING REMARKS

The Periodic Table presented in this work does not arise from group theory alone. Group theory is a branch of mathematics that requires, for its application to a specific domain of physics and chemistry, simultaneous use of physico-chemical theories and/or models. The reader should appreciate, for the $SO(4, 2)\otimes SU(2)$ approach to the Periodic Table, the importance of atomic theory based on quantum mechanics used in conjunction with perturbation and variation methods.

To date, the $SO(4, 2)\otimes SU(2)$ approach to the Periodic Chart has been restricted to qualitative but nonetheless interesting aspects. The KGR program suggested in this work could open an avenue for future investigations beyond qualitative considerations and perhaps make it possible to establish empirical laws for the observed regularities among chemical elements. This might also prove useful for predicting properties of new elements.

Another interest of the $SO(4, 2)\otimes SU(2)$ approach concerns molecules. In this vein, we should mention the work by Hefferlin and Zhuvikin and their collaborators [34]. The starting point of their work is to consider the direct product $[SO(4, 2)\otimes SU(2)]^{\otimes N}$ for describing an N-atom molecule. Preliminary successful tests have been achieved by comparing calculated observables (in a second quantized form adapted to subgroups of $[SO(4, 2)\otimes SU(2)]^{\otimes N}$) with experimental data when available [34].

## 6. ACKNOWLEDGMENTS

I would like to thank the Wiener family and the organizers of the Second Harry Wiener International Memorial Conference, especially Dennis Rouvray (Conference President), Bruce King (Conference Secretary) and Michael Wingham (Conference Vice-President) for making possible this beautiful interdisciplinary meeting where part of this work was presented. I am indebted to Michael Cox for his helpful editorial assistance and suggestions with the final polish. My thanks are also due to Mariasusai Antony and Itsvan Berkes for information concerning superheavy elements.

[4] M. Kibler and P. Winternitz, *Phys. Lett. A*, **147**, 338 (1990).

[5] A.O. Barut and H. Kleinert, *Phys. Rev.*, **156**, 1541 (1967); **157**, 1180 (1967); **160**, 1149 (1967).

[6] I.A. Malkin and V.I. Man'ko, *Pis'ma Zh. Eksp. Teor. Fiz.*, **2**, 230 (1965) [*Sov. Phys., JETP Lett.*, **2**, 146 (1965)].

[7] A.O. Barut and G. Bornzin, *J. Math. Phys.*, **11**, 2938 (1970); **12**, 841 (1971).

[8] C.E. Wulfman, in *Group Theory and Its Applications*, E.M. Loebl, Ed., Vol. II, Academic, New York, 1971; in *Recent Advances in Group Theory and Their Application to Spectroscopy*, J.C. Donini, Ed., Plenum, New York, 1979.

[9] M. Kibler and T. Négadi, *Lett. Nuovo Cimento*, **37**, 225 (1983); *J. Phys. A : Math. Gen.*, **16**, 4265 (1983); *Phys. Rev. A*, **29**, 2891 (1984); *Folia Chim. Theor. Latina*, **12**, 117 (1984); see also: M. Kibler and G. Grenet, *J. Math. Phys.*, **21**, 422 (1980); M. Kibler and T. Négadi, *J. Phys. A : Math. Gen.*, **24**, 5283 (1991).

[10] D. Lambert and M. Kibler, *J. Phys. A: Math. Gen.*, **21**, 307 (1988); M. Kibler, in *Symmetries in Science IX*, B. Gruber and M. Ramek, Eds., Plenum, New York, 1997.

[11] M. Kibler and P. Winternitz, *J. Phys. A: Math. Gen.*, **21**, 1787 (1988).

[12] Yu.B. Rumer and A.I. Fet, *Teor. i Mat. Fiz.*, **9**, 203 (1971) [English translation: *Theor. Math. Phys.*, **9**, 1081 (1972)].

[13] A.O. Barut, in *The Structure of Matter (Proc. Rutherford Centennial Symp., 1971)*, B.G. Wybourne, Ed., University of Canterbury Publications, Bascands, Christchurch, New Zealand, 1972.

[14] H. Odabasi, *Int. J. Quantum Chem., Quantum Chem. Symp.*, **7**, 23 (1973); see also: E.U. Condon and H. Odabasi, *Atomic Structure*, Cambridge University Press, Cambridge, 1980.

[15] A.I. Fet, *Teor. i Mat. Fiz.*, **22**, 323 (1975) [*Theor. Math. Phys.*, **22**, 227 (1975)].

[16] V.M. Byakov, Y.I. Kulakov, Y.B. Rumer and A.I. Fet, Preprint ITEP-26, Moscow (1976); Preprint ITEP-90, Moscow (1976); Preprint ITEP-7, Moscow (1977).

[17] V.G. Konopel'chenko and Yu.B. Rumer, *Usp. Fiz. Nauk*, **129**, 339 (1979) [English translation: *Sov. Phys. Usp.*, **22**, 837 (1979)].

[18] M. Kibler, *J. Mol. Struct. (Theochem)*, **187**, 83 (1989).

[19] L.I. Gurskii, L.I. Komarov and A.M. Solodukhin, *Int. J. Quantum Chem.*, **72**, 499 (1999).

[20] H. Bateman, *J. Lond. Math. Soc.*, **8**, 70 (1908); E. Cunningham, *J. Lond. Math. Soc.*, **8**, 77 (1909); H. Bateman, *J. Lond. Math. Soc.*, **8**, 223 (1909).

[21] O.A. Novaro and K.B. Wolf, *Rev. Mex. Fís.*, **20**, 265 (1971); O.A. Novaro and M. Berrondo, *J. Phys. B: At. Mol. Phys.*, **5**, 1104 (1972); M. Berrondo and O.A. Novaro, *J. Phys. B: At. Mol. Phys.*, **6**, 761 (1973); O. Novaro, *Int. J. Quantum Chem., Quantum Chem. Symp.*, **7**, 53 (1973); see also: O. Novaro, *J. Mol. Struct. (Theochem)*, **199**, 103 (1989).

[22] V.N. Ostrovsky, *J. Phys. B: At. Mol. Phys.*, **14**, 4425 (1981); see also: Yu.N. Demkov and V.N. Ostrovsky, *Zh. Eksp. Teor. Fiz.*, **60**, 2011 (1971) [English translation: *Sov. Phys., JETP,* **33**, 1083 (1971)]; *Zh. Eksp. Teor. Fiz.*, **62**, 125 (1971); Errata, **63**, 2376 (1972) [English translation: *Sov. Phys., JETP,* **35**, 66 (1972)].

[23] Y. Kitagawara and A.O. Barut, *Phys. Rev. B*, **16**, 3305 (1983); *J. Phys. B: At. Mol. Phys.*, **16**, 3305 (1983); **17**, 4251 (1984).

[24] T. Négadi and M. Kibler, *J. Phys. A: Math. Gen.*, **25**, L157 (1992); *Int. J. Quantum Chem.*, **57**, 53 (1996).

[25] D. Neubert, *Z. Naturforsch., a*, **25**, 210 (1970).

[26] H.H. Dash, *Int. J. Quantum Chem., Quantum Chem. Symp.*, **3**, 335 (1969).

[27] E.R. Scerri, in *The Periodic Table: Into the 21st Century*, D.H. Rouvray and R.B. King, Eds., Research Studies Press, Baldock, United Kingdom, 2004.

[28] M.S. Antony, *Tableau périodique des éléments*, Institut de Recherches Subatomiques, Strasbourg, 2004; Yu.Ts. Oganessian et al., *Phys. Rev. C*, **69**, 021601 (2004); **69**, 029902 (2004).

---

[29] F. Trombe, J. Loriers, F. Gaume-Mahn and C. Henry La Blanchetais, in *Nouveau traité de chimie minérale VII*; P. Pascal, Ed., Masson and Cie, Paris, 1960; X. Oudet, *J. Physique*, **40**, C5-395 (1979); *Ann. Fondation L. de Broglie*, **6**, 127 (1981); X. Oudet and G. Lochak, *J. Mag. Magnetic Materials*, **65**, 99 (1987).

[30] E. Madelung, *Die mathematischen Hilfsmittel des Physikers*, Springer, Berlin, 1936; V.M. Klechkovskii, *J. Exp. Theor. Phys.*, **41**, 465 (1961); S.A. Goudsmit and P.I. Richards, *Proc. Natl. Acad. Sci. U.S.A.*, **51**, 664 (1964).

[31] M. Laing, *Educ. Chem.*, **36**, 160 (1999).

[32] J. Katriel and C.K. Jørgensen, *Chem. Phys. Lett.*, **87**, 315 (1982); D. Purdela, *Int. J. Quantum Chem.*, **34**, 107 (1988).

[33] N. Bohr, *Z. Phys.*, **9**, 1 (1922).

[34] G.V. Zhuvikin and R. Hefferlin, *Vestnik Leningradskovo Universiteta*, **16**, 10 (1983); R.A. Hefferlin, G.V. Zhuvikin, K.E. Caviness and P.J. Duerksen, *J. Quantum Spectrosc. Radiat. Transfer*, **32**, 257 (1984); G.V. Zhuvikin and R.A. Hefferlin, in *Anales de Física, Monografías*, M.A. del Olmo, M. Santander and J.M. Guilarte, Eds., Real Sociedad Española de Física, Madrid, 1993; G.V. Zhuvikin and R. Hefferlin, *Symmetry Principles for Periodic Systems of Molecules*, Joint Report SC/SPBU#1, 1994; C.M. Carlson, R.A. Hefferlin and G.V. Zhuvikin, *Analysis of Group Theoretical Periodic Systems of Molecules Using Tabulated Data*, Joint Report SC/SPBU#2, 1995.